\documentclass[aps,prd,nofootinbib,twocolumn,superscriptaddress,preprintnumbers]{revtex4}

\usepackage{amsmath,amssymb,amsthm,color,slashed}
\usepackage{graphicx}
\usepackage{dcolumn}
\usepackage{bm}
\usepackage{siunitx}

\begin{document}

\preprint{CALT-TH-2015-045, IPMU15-0145, UCB-PTH-15/07}

\title{Weak Gravity Conjecture in AdS/CFT}

\author{Yu Nakayama}

\affiliation{Walter Burke Institute for Theoretical Physics,
 California Institute of Technology, Pasadena, CA 91125, USA}
\affiliation{Kavli Institute for the Physics and Mathematics of the 
 Universe (WPI), UTIAS, The University of Tokyo, Kashiwa, 
 Chiba 277-8583, Japan}

\author{Yasunori Nomura}

\affiliation{Berkeley Center for Theoretical Physics, Department of Physics, 
 University of California, Berkeley, CA 94720, USA}
\affiliation{Theoretical Physics Group, Lawrence Berkeley National Laboratory,
 Berkeley, CA 94720, USA}

\begin{abstract}
We study implications of the weak gravity conjecture in the AdS/CFT 
correspondence. Unlike in Minkowski spacetime, AdS spacetime has a physical 
length scale, so that the conjecture must be generalized with an additional 
parameter.  We discuss possible generalizations and translate them into 
the language of dual CFTs, which take the form of inequalities involving 
the dimension and charge of an operator as well as the current and 
energy-momentum tensor central charges.  We then test these inequalities 
against various CFTs to see if they are universally obeyed by all the 
CFTs.  We find that certain CFTs, such as supersymmetric QCDs, do not 
satisfy them even in the large $N$ limit.  This does not contradict 
the conjecture in AdS spacetime because the theories violating them 
are either unlikely or unclear to have weakly coupled gravitational 
descriptions, but it suggests that the CFT inequalities obtained here 
by naive translations do not apply beyond the regime in which weakly 
coupled gravitational descriptions are available.
\end{abstract}

\maketitle

\section{Introduction}
\label{sec:intro}

There are many folk theorems that are believed to hold in quantum 
gravity.  Some are qualitative such as the non-existence of continuous 
global symmetries, suggested by the physics of black holes as well 
as perturbative string theory.  Others are more quantitative, which 
include the weak gravity conjecture~\cite{ArkaniHamed:2006dz}.  These 
more quantitative theorems, however, generally have weaker foundations 
and their precise meanings are obscured beyond the semiclassical limit. 
For recent discussions on the weak gravity conjecture, see e.g.\ 
Refs.~\cite{Cheung:2014vva,Montero:2015ofa,Brown:2015iha,Heidenreich:2015wga}.

Since AdS/CFT duality~\cite{Maldacena:1997re,Gubser:1998bc} provides 
a nonperturbative definition of quantum gravity, it is natural to 
explore how the folk theorems in quantum gravity may be realized in 
this framework.  Ideally, a folk theorem can be translated into a 
universal statement in CFTs which may be tested, at least under some 
circumstances.  Alternatively, one might find that such a universal 
statement is not possible, in which case one would learn that the 
theorem arises as a property that manifests itself only in a certain 
(weakly coupled gravitational) limit of the theory.

Motivated by these considerations, in this article we study the weak 
gravity conjecture in AdS/CFT.  Since the original conjecture was 
formulated in asymptotically Minkowski spacetime, we first discuss 
possible generalizations in AdS spacetime (in Section~\ref{sec:AdS}). 
Then, we translate the statements into the language of CFTs, all of 
which take the form that there must be an operator whose coupling to 
the energy-momentum tensor is smaller than that to the conserved current 
(in Section~\ref{sec:CFT}).  Finally, we test these statements against 
known CFTs (in Section~\ref{sec:test}).  We find that the statements 
as formulated here do not apply universally to all the CFTs.  On 
the other hand, all the theories that do not satisfy them are those 
that are believed not to have weakly coupled gravitational descriptions 
or unclear to have such descriptions.  It is, therefore, still consistent 
to postulate that the weak gravity bounds discussed here hold in 
asymptotically AdS spacetime.  An alternative possibility is that 
there are some modified expressions that apply universally and reduce 
to the bounds discussed here when there are weakly coupled gravitational 
descriptions.  This is discussed in Section~\ref{sec:discuss}.

\section{Weak Gravity Conjecture in Minkowski Spacetime}
\label{sec:Minkowski}

Consider Einstein-Maxwell theory in $D$-dimensional (asymptotically-)flat 
Minkowski spacetime
\begin{equation}
  S = \int d^D x \sqrt{-g} \left( \frac{R}{2\kappa_D^2} 
    - \frac{1}{4e^2} F_{MN} F^{MN} + \text{matter} \right),
\label{eq:Einstein-Maxwell}
\end{equation}
where $\kappa_D^{2} = M_{\rm Pl}^{2-D}$ is the $D$-dimensional Newton 
constant.  The weak gravity conjecture states that a low energy effective 
theory of a consistent theory of quantum gravity must contain a particle 
with the mass $m$ and charge $q$ satisfying%
\footnote{There are two versions of the conjecture discussed in 
 Ref.~\cite{ArkaniHamed:2006dz}.  In this paper we focus on the weaker 
 (more conservative) version.}
\begin{equation}
  \frac{m^2}{q^2} \le C_D e^2 \kappa_D^{-2}.
\label{eq:wgc-flat}
\end{equation}
Here, the coefficient $C_D$ is determined such that the inequality 
is saturated by the extremal Reissner-Nordstr\"{o}m (RN) black hole of 
mass $m$ and charge $q$.  (In the normalization of $q e$ we will adopt 
later, $C_D = (D-2)/(D-3)$.)

An alternative, and essentially equivalent, formulation of the conjecture 
is given by the statement that extremal (non-BPS) RN black holes 
must be unstable, at least marginally.  A connection between the 
two formulations is the following.  Imagine that the weak gravity 
conjecture were violated.  Then the particle with the smallest $m/q$ 
has $\kappa_D^2 m^2 > C_D e^2 q^2$, so that gravitational attraction 
between two such particles is stronger than the gauge repulsion.  This 
implies that we can form Kepler bound states composed of any number $n$ 
of these particles, which are all absolutely stable and become extremal 
in the $n \to \infty$ limit.  On the other hand, if there exists 
a particle with $m < \sqrt{C_D} q e/\kappa_D$, then extremal RN black 
holes can decay, except possibly for ``quantum'' ones with charges 
smaller than $q$, where we have assumed $q \,\slashed{\gg}\, 1$.

In this paper we adopt the latter formulation, based on extremal 
RN black holes, and discuss how it may be generalized in AdS spacetime. 
We also see how the generalized conjecture may be interpreted in dual 
CFTs, using the AdS/CFT correspondence.

\section{Weak Gravity Conjecture in AdS Spacetime}
\label{sec:AdS}

How can we extend the weak gravity conjecture to asymptotically AdS 
spacetime?  The answer is not obvious because of the following facts: 
(i) AdS spacetime can be regarded as a finite box, preventing Hawking 
radiation from escaping to ``infinity''; (ii) Physical properties of 
black holes change when their size becomes larger than the AdS scale 
(making the $n \to \infty$ limit we took in the previous section less 
convincing); (iii) Unlike in Minkowski spacetime, there is no no-hair 
theorem in AdS spacetime, making it possible for a black hole to decay 
by a process that does not have a direct analogue in Minkowski spacetime.

Given these facts, in this paper we formulate our conjecture(s) in the 
following steps.%
\footnote{Related but different conjectures were discussed in 
 Refs.~\cite{Huang:2006hc,Kovtun:2008kx}.}
We first consider the requirement that small extremal AdS-RN black holes 
must be able to decay by a process that is also available in Minkowski 
spacetime.  In particular, we require that there is a particle in the 
AdS theory to which small extremal AdS-RN black holes can decay.  We 
call this condition the {\it simple kinematic conjecture}, and discuss 
its formulation in dual CFTs.

We next consider the condition that small extremal AdS-RN black holes 
decay by a dynamical process that is available (only) in AdS spacetime. 
In particular, we consider that the decay occurs through superradiant 
instability discussed in Refs.~\cite{Bhattacharyya:2010yg,Dias:2010ma}. 
We find that this gives a condition weaker than that of the simple 
kinematic conjecture, and call it the {\it dynamical conjecture}. 
The difference between the simple kinematic and dynamical conjectures 
is purely AdS in nature---both these conjectures reduce to the 
Minkowski one in the appropriate large AdS radius limit.

We finally discuss possible additional constraints coming from large 
extremal AdS-RN black holes.  We find that as long as either of the 
above conjectures is satisfied, a large extremal AdS-RN black hole 
can always have a microscopic ``decay'' process.  Namely, a process 
in which a larger black hole is converted into a smaller one and the 
light quantum is always kinematically allowed.  It is possible that 
this is indeed enough for the consistency of the theory.

On the other hand, in AdS spacetime the above process does not lead 
to a real decay of a large black hole because the finite-box nature of 
AdS makes a large black hole be in thermal equilibrium with the ambient 
space.  To make the large black hole really unstable, we need to have 
a different process.  In Ref.~\cite{Denef:2009tp}, it was advocated 
that this may in fact be the case---large extremal AdS-RN black holes 
have instabilities associated with the presence of a superconducting 
phase in strongly coupled dual CFTs.  While we do not have a better 
argument for this conjecture than the authors of Ref.~\cite{Denef:2009tp}, 
we also discuss it for completeness.

\section{CFT Formulation}
\label{sec:CFT}

We now formulate our conjectures using the language of dual CFTs. 
Below, we focus on the case with $D=5$, but the extension to other 
dimensions is straightforward.

In dual $d=4$ CFTs, a conserved current $J_\mu$ and the energy-momentum 
tensor $T_{\mu\nu}$ have the two-point functions
\begin{align}
  \langle J_{\mu}(x) J_\nu(0) \rangle 
    &= \frac{C_V}{x^6} I_{\mu\nu}(x),
\label{eq:def-CV}\\
  \langle T_{\mu\nu}(x) T_{\rho\sigma}(0) \rangle 
    &= \frac{C_T}{x^8} I_{\mu\nu,\rho\sigma}(x),
\label{eq:def-CT}
\end{align}
where $I_{\mu\nu}(x) = \delta_{\mu\nu} -x_\mu x_\nu/x^2$ and 
$I_{\mu\nu,\rho\sigma}(x) = (I_{\mu\rho}(x)I_{\nu\sigma}(x) + 
I_{\mu\sigma}(x)I_{\nu\rho}(x))/2 - \delta_{\mu\nu} \delta_{\rho\sigma}/4$. 
Crudely speaking, $C_T$ counts the number of massless degrees of freedom 
in the CFT, while $C_V$ counts the number of massless {\it charged} 
degrees of freedom in the CFT. (When the current $J_\mu$ is gauged, 
the leading-order beta function is proportional to $C_V$.)  For 
our explicit normalization convention for these quantities, see 
Appendix~\ref{app:norm}.

The AdS/CFT correspondence states that $C_V$ and $C_T$ are related to 
the kinetic terms of the bulk fields in AdS spacetime
\begin{equation}
  S =  \int\! d^5 x \sqrt{-g} 
    \left( \frac{1}{2\kappa_5^2} \left(R +\frac{12}{L^2}\right) 
    - \frac{1}{4e^2} F_{MN}F^{MN} + \cdots \right),
\label{eq:kinetic}
\end{equation}
as
\begin{equation}
  C_V = \frac{6 L}{\pi^2} e^{-2},
\qquad
  C_T = \frac{40 L^3}{\pi^2} \kappa_5^{-2},
\label{eq:CV-CT}
\end{equation}
where $L$ is the AdS radius.  The existence of a bulk field of mass $m$ 
implies that of a CFT operator of scaling dimension
\begin{equation}
  \Delta = L m + O(1),
\label{eq:mass}
\end{equation}
where $O(1)$ corrections depend on the spin and detailed couplings, and 
we will discuss it only when necessary.  It is natural to focus on $\Delta$ 
rather than $m$, since it corresponds to the conserved global energy in 
AdS spacetime.  A (minimally-coupled) scalar field in AdS spacetime must 
satisfy the Breitenlohner-Freedman bound
\begin{equation}
  m^2 L^2 \ge -4.
\label{eq:BF-bound}
\end{equation}
Note that a small negative mass-squared is allowed without causing 
an instability.

\subsection{Simple kinematic conjecture}
\label{subsec:kinematic}

Let us first consider the simple kinematic bound coming from the 
requirement that there exists a particle that has a smaller ratio of 
the AdS energy $\Delta$ to the charge $q$ than that of small extremal 
AdS-RN black holes (which have the horizon sizes smaller than the 
AdS radius $L$).  As summarized in Appendix~\ref{app:AdS-RN}, in AdS 
spacetime the mass-to-charge ratio, $M/Q$, of a small extremal black 
hole depends on the size of the black hole
\begin{equation}
  \frac{M^2}{Q^2} = \frac{3 e^2}{2 \kappa^2} h(x),
\label{eq:M-to-Q}
\end{equation}
where $h(x) = (3 x^2/4) (\sqrt{1+x}-1)^{-2} (2\sqrt{1+x}+1)^{-1}$, 
and $x = 2 M \kappa_5^2/L^2$ ($0 < x \lesssim 1$).  Since $h(x)$ is 
a monotonically increasing function, however, requiring the bound 
for the smallest black hole, i.e.\ in the $x \rightarrow 0$ limit,%
\footnote{This limit must be taken such that the size of the black hole 
 is still larger than the 5D Planck scale.  In the CFT language, if we 
 have a (5D) Planck-sized black hole, $\Delta \sim L \kappa_{5}^{-2/3} 
 \sim C_T^{1/3}$.  In comparison, we have $\Delta \gtrsim C_T$ for large 
 black holes.  Note that the 5D Planck scale is the largest conceivable 
 cutoff for the 5D theory, but there can be lower scales such as the 
 Kaluza-Klein or string scales.  In fact, Ref.~\cite{ArkaniHamed:2006dz} 
 argues that this must be the case, based on an analysis of 5D AdS 
 spacetime cut off by a ``UV brane.''}
ensures that all heavier black holes satisfy the corresponding 
bounds.

This leads to the condition that in 5D AdS spacetime there must be a 
particle whose AdS energy $E$ and charge $q$ satisfy
\begin{equation}
  \frac{E^2}{q^2} \le \frac{3}{2} e^2 \kappa^{-2}_5.
\label{eq:kin-1}
\end{equation}
Using Eq.~(\ref{eq:CV-CT}) and $\Delta = LE$, we can write this in terms 
of the CFT data%
\footnote{Whether this inequality is satisfied or not is related to 
 a certain convexity of the CFT operator spectrum in the large spin 
 limit~\cite{Komargodski:2012ek}. We thank Jo\~{a}o Penedones for 
 bringing this to our attention.}
\begin{equation}
  \frac{\Delta^2}{q^2} \le \frac{9}{40} \frac{C_T}{C_V}.
\label{eq:naive}
\end{equation}
This condition, by itself, does not tell us where the state exists, but 
it is natural to expect that it must be below the mass of the lightest 
5D AdS-RN black hole.

The mass of the lightest 5D AdS-RN black hole depends on the size of the 
extra dimensions beyond 5D AdS we consider.  It is not known how small 
the extra dimensions can be made in general, but it is possible that there 
is a lower bound on their size.  For example, Ref.~\cite{ArkaniHamed:2006dz} 
argues that the volume of the extra dimensional space $X$ must satisfy 
$(V_X/l_{\rm s}^5) \gtrsim g_{\rm s} (R/l_{\rm s})$, where $l_{\rm s}$ 
and $g_{\rm s}$ are the string length and coupling, respectively.  Assuming 
that $X$ has only one length scale, this implies that the state satisfying 
Eq.~(\ref{eq:naive}) must exist below $\Delta \sim C_T^{3/5}$.

Since black holes in the $x \rightarrow 0$ limit behave similarly to 
those in Minkowski spacetime, we expect that the condition discussed here 
is reduced to the original Minkowski bound when we take $L \to \infty$ 
(with the fixed Planck scale as well as any other scales).  Indeed, using 
Eqs.~(\ref{eq:CV-CT},~\ref{eq:mass}), we find that Eq.~(\ref{eq:naive}) 
yields Eq.~(\ref{eq:wgc-flat}) in the appropriate limit.

\subsection{Dynamical conjecture}
\label{subsec:dynamical}

In general, the stability condition for a system in AdS spacetime is 
different from that in Minkowski spacetime.  In particular, since there 
is no no-hair theorem in AdS spacetime, extremal AdS-RN black holes may 
have dynamical instabilities involving classical condensates, which are 
not available in Minkowski spacetime.  Indeed, it is known that in the 
presence of a minimally coupled charged scalar field, extremal AdS-RN 
black holes may be unstable against scalar hair formation.  For small 
extremal AdS-RN black holes, this instability can be interpreted as 
a superradiant instability.

According to Refs.~\cite{Bhattacharyya:2010yg,Dias:2010ma}, the 
superradiant instability for small extremal black holes occurs when 
there is a minimally coupled charged scalar field in the bulk satisfying 
the condition (for the $r_+/L \to 0$ limit):
\begin{equation}
  L^2 m^2 - \frac{3}{2} e^2 \kappa_5^{-2} L^2 q^2 \le -4.
\label{eq:superrad}
\end{equation}
In terms of the CFT data, this leads to
\begin{equation}
  \frac{(\Delta-2)^2}{q^2} \le \frac{9}{40} \frac{C_T}{C_V},
\label{eq:wgcads}
\end{equation}
where $\Delta$ is the dimension of the CFT operator corresponding to the 
charged scalar field in the bulk.  Note that the bound on $\Delta$ is 
shifted by two units compared with that in Eq.~(\ref{eq:naive}).  This 
is because the condensation effect can make the AdS energy per charge 
lower than that of the collection of quanta.  In fact, in the range 
allowed by unitarity, $\Delta \geq 1$, the bound in Eq.~(\ref{eq:wgcads}) 
is weaker than that in Eq.~(\ref{eq:naive}).

In the appropriate Minkowski limit (sending $L \to \infty$ while keeping 
$m$), Eq.~(\ref{eq:wgcads}) is also reduced to the Minkowski bound 
in Eq.~(\ref{eq:wgc-flat}).  This implies that the difference between 
the two bounds in Eqs.~(\ref{eq:naive}) and (\ref{eq:wgcads}) is purely 
AdS in nature---it is important only for low $\Delta$.

We note that, unlike the corresponding objects in Minkowski spacetime, 
extremal AdS-RN black holes do not saturate the BPS bound (i.e.\ they 
cannot be supersymmetric), except in the limit $r_+/L \to 0$.  (See 
Appendix~\ref{app:AdS-RN}.) The decay processes described above, 
therefore, may occur non-marginally even in theories with supersymmetry.

\subsection{Large black holes}
\label{subsec:large-BH}

In AdS spacetime, we have large extremal AdS-RN black holes ($r_+ > L$), 
which do not possess a simple flat spacetime limit.  While the weak gravity 
bound in Minkowski spacetime does not directly lead to the conclusion 
that these black holes must be unstable, it is interesting to see what 
bounds on CFTs can be obtained by requiring that they are indeed unstable. 
In fact, the idea that the planar extremal AdS-RN black branes (which 
are equivalent to AdS-RN black holes in the $r_+/L \to \infty$ limit) 
should be unstable was advocated in Ref.~\cite{Denef:2009tp}, in relation 
to the presence of a superconducting phase in strongly coupled CFTs.

The instability condition on an AdS-RN black hole with respect to the 
formation of (minimally coupled) scalar hair condensation depends on 
the size of the black hole $r_+$.  While the general condition can be 
found in Ref.~\cite{Dias:2010ma}, here we quote only two representative 
cases.  We expect that the true bound is obtained by the union of the 
conditions for all values of $r_+ \gtrsim L$.

In the limit of a planar extremal AdS-RN black brane (i.e.\ $r_+/L \to 
\infty$), the horizon topology becomes $\mbox{AdS}_2 \times R^3$, and 
the instability appears when the effective mass of a charged field near 
the horizon becomes below the $\mbox{AdS}_2$ Breitenlohner-Freedman 
bound.  In our normalization, we find (\cite{Denef:2009tp} for $D=4$ 
and \cite{Gubser:2009qm} for $D=5$)
\begin{equation}
  \frac{3}{2} \frac{(\Delta-1)(\Delta-3)}{q^2} 
    \le \frac{9}{40} \frac{C_T}{C_V}.
\label{eq:large-1}
\end{equation}
On the other hand, for an ``intermediate'' extremal AdS-RN black hole 
(i.e.\ $r_+ \sim L$ or $\Delta_{\rm BH} \sim C_T$), the condition that 
it must be unstable gives (for $r_+ = L$)
\begin{equation}
  \frac{4}{3} \frac{(\Delta-2)^2}{q^2} 
    \le \frac{9}{40} \frac{C_T}{C_V}.
\label{eq:large-2}
\end{equation}
The shift in $\Delta$ is the same as in Eq.~(\ref{eq:wgcads}), but we 
have an additional factor of $4/3$ in the left-hand side.

The conditions in Eqs.~(\ref{eq:large-1},~\ref{eq:large-2}) give stronger 
bounds than the original weak gravity bound, Eq.~(\ref{eq:wgc-flat}), in 
the naive flat-space limit $\Delta \gg 1$.  This, however, does not mean 
the existence of a stronger bound than Eq.~(\ref{eq:wgc-flat}) in Minkowski 
spacetime.  In the true Minkowski limit, large black holes considered 
here disappear from the spectrum, and so do the corresponding bounds.

\section{Testing with Examples}
\label{sec:test}

In this section, we study if the bounds discussed in the previous section 
are indeed satisfied in various known CFTs.  Since our conjectures are 
about ``generic'' CFTs that have weakly coupled gravitational descriptions, 
and these theories are not well understood, we need to ``test'' them 
against theories in our hands, which are not necessarily in a class to 
which the conjectures must apply.  Nevertheless, we find some interesting 
lesson---all the theories that we find do not satisfy the bounds are those 
that are believed not to have weakly coupled gravitational descriptions 
(or unclear to have such descriptions).  In particular, we find that 
supersymmetric theories that have weakly coupled gravitational descriptions 
(although in 10D) do satisfy the bounds.

\subsection{Known AdS/CFT with supersymmetry}
\label{subsec:known}

We first study if our conjectures are satisfied in known examples of 
the AdS/CFT correspondence.  Since these theories have weakly coupled 
gravitational descriptions in 10D, our analysis in the previous section 
need not a priori apply.  Moreover, their 10D bulk descriptions possess 
high supersymmetries that relate gravity with $U(1)$ gauge forces, 
reducing the significance of the conjectures in some cases.  Nevertheless, 
we find it nontrivial that all these theories satisfy the bounds applied 
naively, especially given that not all the CFTs satisfy them as we will 
see in later subsections.

To be specific, we focus on type~IIB string theory compactified on 
$\mbox{AdS}_5 \times Y_{p,q}$ with the coprime numbers $p>q$, which 
have weakly coupled supergravity descriptions with the second order bulk 
actions.  In these theories, the Kaluza-Klein reduction is consistent 
and most 5D asymptotic AdS solutions (including black holes) can be 
uplifted to 10D solutions~\cite{Gauntlett:2007ma} despite the intrinsic 
10D nature of these theories.  The compact spaces $Y_{p,q}$ are nontrivial 
examples of Sasaki-Einstein five folds, whose explicit construction 
can be found in Ref.~\cite{Gauntlett:2004yd}.

The resulting dual CFTs preserve $\mathcal{N}=1$ superconformal symmetry 
in 4D with $U(1)_R$ symmetry.  The central charges for the energy-momentum 
tensor and the $R$ current can be computed both from the gravitational 
and field theory points of view, giving~\cite{Barnes:2005bw}
\begin{align}
  C_T &= \frac{10 N^2}{\pi V_{p,q}},
\\
  C_R &= \frac{ N^2}{\pi V_{p,q}},
\end{align}
where $N$ and $V_{p,q}$ are the number of branes and the volume of 
$Y_{p,q}$, respectively (which we will not use).  These theories also 
have $U(1)_F \times U(1)_B \times SU(2)$ global symmetries.  As long as 
we have a scalar chiral operator, the superconformal $R$ charge always 
saturates the simple kinematic bound in Section~\ref{subsec:kinematic} 
(and thus satisfies the weaker bound in Section~\ref{subsec:dynamical}). 
We are therefore more interested in other global symmetries.

Let us first consider the $U(1)_F$ symmetry.  In the theories under 
consideration, there are three types of chiral primary operators with 
their $U(1)_F$ charges given in terms of the $R$ charges as
\begin{align}
  q_F(O_1) &= y_1 R(O_1), \\
  q_F(O_2) &= -y_2 R(O_2), \\
  q_F(O_3) &= -\frac{1}{2}(y_1 + y_2) R(O_3),
\end{align}
where
\begin{align}
  y_1 &= \frac{1}{4p}(2p-3q-\sqrt{4p^2-3q^2}), \\
  y_2 &= \frac{1}{4p}(2p+3q-\sqrt{4p^2-3q^2}).
\end{align}
Since these are scalar chiral primary operators, they satisfy $\Delta(O_i) 
= (3/2)R(O_i)$.  The AdS/CFT as well as direct field theory computations 
give the central charge for $U(1)_F$ as
\begin{equation}
  C_F = \frac{N^2}{8 \pi V_{p,q}} 
    \frac{\sqrt{4p^2-3q^2}}{p^2} (2p-\sqrt{4p^2-3q^2}).
\label{eq:C_F}
\end{equation}
Using these formulae, we can calculate the ratios $\Delta^2/q_F^2$ for 
$O_{1,2,3}$.  We find that the simple kinematic bound
\begin{align}
  \frac{\Delta^2}{q_F^2} \le \frac{9}{40} \frac{C_T}{C_F},
\label{eq:bound-U1F}
\end{align}
is always satisfied by $O_1$ and $O_2$  (but not necessarily $O_3$). 
Note that in order to be consistent with the weak gravity conjecture, 
we only need one operator (e.g.\ $O_1$ here) that satisfies the bound. 
The most stringent case is the $p\gg q$ limit, but we still have a factor 
of $3$ margin there.

As far as we have checked, in all known examples of the AdS/CFT 
correspondence with weakly coupled gravity descriptions, the simple 
kinetic bound in Section~\ref{subsec:kinematic} (and thus also the 
dynamical bound in Section~\ref{subsec:dynamical}) is satisfied for 
the $R$ symmetries and Abelian flavor symmetries.  The further such 
examples include $\mbox{AdS}_5 \times L_{p,q,r}$ compactification of 
type~IIB string theory~\cite{Butti:2005sw}.  We find this nontrivial.

As for the baryonic symmetry, the situation is less clear.  In the examples 
considered, the lightest object charged under the baryonic symmetry has 
$\Delta \sim N \sim C_T^{1/2}$, so that it is heavier than the 5D Planck 
scale, $\Delta \sim L \kappa_{5}^{-2/3} \sim C_T^{1/3}$.  This, however, 
may not mean a violation of the bound if the size of extra dimensions 
is necessarily larger than the (effective) 5D Planck scale; see discussions 
after Eq.~(\ref{eq:naive}).

Let us now turn to the bound coming from large black holes, discussed 
in Section~\ref{subsec:large-BH}.  Recall that this bound is related, 
in the limit $r_+/L \to \infty$, to the (in)stability of planer 
extremal AdS-RN black branes, since in this limit the horizon can 
be approximated by a plane with $\mathbf{R}^3$ topology.  In fact, 
there had been some interests in the stability of these objects in 
string compactification~\cite{Denef:2009tp,Gubser:2009qm,Klebanov:2010tj}. 
The motivation there was mainly applications to condensed matter physics, 
in which the instability of these objects corresponds to the instability 
of zero temperature CFTs under the introduction of chemical potentials. 
In Section~\ref{subsec:large-BH}, we discussed a possible instability 
due to a scalar hair formation.  In the dual CFT language, this 
corresponds to an instability of the system due to a scalar condensate, 
leading to a superfluidity or superconductivity phase transition.

In all the examples studied in Refs.~\cite{Denef:2009tp,Gubser:2009qm,%
Klebanov:2010tj}, the extremal AdS-RN black branes are indeed (marginally) 
unstable due to such scalar condensates.  References~\cite{Denef:2009tp,%
Gubser:2009qm} studied (mainly) $R$-charged extremal AdS-RN black 
branes in which $R$-charged scalar fields, typically dual to 
chiral primary operators in the CFTs, trigger the instability.  In 
Ref.~\cite{Klebanov:2010tj}, a more intricate situation with baryon 
charges was studied and the system was still marginally (un)stable.%
\footnote{Strictly speaking, what they obtained by their tree-level 
 computations is that the potential vanishes.  It is, however, 
 conjectured that higher order corrections make the system unstable. 
 We thank Igor Klebanov for discussions.}
These authors interpreted this observation as a manifestation of the 
weak gravity conjecture applied to extremal AdS-RN branes.  In our 
viewpoint, these examples suggest that extremal AdS-RN black holes 
are unstable in the large black hole limit.  Correspondingly, in these 
dual CFTs, there exists an operator that (marginally) satisfies the 
bound such as Eqs.~(\ref{eq:large-1},~\ref{eq:large-2}).

\subsection{Free theories}
\label{subsec:free}

We now study if our bounds, as formulated in Eqs.~(\ref{eq:naive}), 
(\ref{eq:wgcads}), (\ref{eq:large-1}), and (\ref{eq:large-2}), can 
be universally valid for all the CFTs regardless of the existence of 
a weakly coupled gravitational picture.

For this purpose, let us consider free field theories.  We find that 
the naive bound in Eq.~(\ref{eq:naive}) cannot be universal.  Take 
a free complex scalar with a $U(1)$ global symmetry.  This theory has 
an operator (free complex scalar itself) with $\Delta = 1$.  Normalizing 
the charge of this scalar to be unity, $q=1$, we find that $C_T/C_V = 8/3$. 
The bound in Eq.~(\ref{eq:naive}) then leads to $1 \le 3/5$, which is 
clearly not satisfied.  The existence of other operators does not help, 
since they all have $|\Delta/q| \geq 1$.  A similar conclusion is also 
obtained for a free fermion.

The situation is different for the dynamical conjecture in 
Eq.~(\ref{eq:wgcads}), which gives a weaker bound.  This bound is 
satisfied by a free scalar and a free fermion due to the shift in 
the left-hand side.  It is trivially satisfied for a charged free 
scalar $\phi$ because of the existence of the $\phi^2$ operator, 
which has $(\Delta,q) = (2,2)$.  For a free fermion $\psi$, we have 
a $(\Delta,q) = (3,2)$ scalar operator (i.e.\ $\psi^2$), and since 
the theory has $C_T/C_V =2$, the bound is satisfied.  Therefore, 
at this point, the bound in Eq.~(\ref{eq:wgcads}) still has a chance 
to be universal.

Finally, we discuss the bounds in Eqs.~(\ref{eq:large-1},~\ref{eq:large-2}), 
arising from considerations of large black holes.  These bounds are 
also satisfied by free scalars and fermions.  The meaning of this fact, 
however, is not clear.  In a weakly coupled gravitational description, 
we might as well formulate the conjecture in a form more physical from 
the CFT point of view:\ the zero temperature CFTs must be unstable under 
the introduction of a chemical potential.  (This issue was studied in 
Ref.~\cite{Hartnoll:2011pp}.)  In this form, however, we know that a 
free fermion system does not satisfy the conjecture, since it is stable 
under the introduction of large chemical potentials.  This casts some 
doubt on adopting Eqs.~(\ref{eq:large-1},~\ref{eq:large-2}) as the 
conditions applying universally beyond the weakly coupled gravity limit.%
\footnote{We thank Sean Hartnoll for discussions on this point.}

\subsection{Supersymmetric QCDs in the large $N$ limit}
\label{subsec:SQCD}

Consider supersymmetric QCDs with $SU(N_c)$ gauge groups and $N_f$ flavors 
of quarks (i.e.\ $N_f$ $Q$'s and $N_f$ $\bar{Q}$'s in the fundamental 
and anti-fundamental representations of $SU(N_c)$, respectively) in the 
conformal window $\frac{3}{2}N_c \le N_f \le 3N_c$.%
\footnote{We take the Veneziano limit $N_c, N_f \to \infty$ for the 
 purpose of simplifying the formula.}
This theory possesses a $U(1)_B$ symmetry, $Q(+1)$ and $\bar{Q}(-1)$, 
in addition to the $SU(N_f) \times SU(N_f)$ flavor symmetry and the 
$R$-symmetry.  Since the theory has scalar chiral superconformal primary 
operators, the $R$-symmetry automatically satisfies the bounds in 
Eqs.~(\ref{eq:naive},~\ref{eq:wgcads}).  We thus focus on the $U(1)_B$ 
symmetry below.

While the theory is strongly coupled away from the perturbative regime 
$N_f \sim 3 N_c$, one can compute the exact value of the $U(1)_B$-current 
central charge $C_B$ from the supersymmetric formula
\begin{equation}
  C_B = -\frac{9}{4\pi^4}\mathrm{Tr}[R BB] 
  = \frac{9}{4\pi^4}(2 N_f N_c) \frac{N_c}{N_f}.
\label{eq:SQCD:C_B}
\end{equation}
Similarly, the exact value of the energy-momentum tensor central charge 
is~\cite{Anselmi:1997am}
\begin{equation}
  C_T = \frac{5}{2\pi^4} (7 N_c^2 - \frac{9 N_c^4}{N_f^2}),
\label{eq:SQCD:C_T}
\end{equation}
leading to
\begin{equation}
  \frac{C_T}{C_B} = \frac{5}{9} (7 - \frac{9 N_c^2}{N_f^2}).
\label{eq:SQCD:C_T-C_B}
\end{equation}

Because of the gauge invariance, the lightest baryonic charged operator 
is $\epsilon QQQQ...$ (with $N_c$ $Q$'s), which has
\begin{equation}
  q_B = N_c,
\qquad
  \Delta = \frac{3}{2}N_c (1-\frac{N_c}{N_f}).
\label{eq:baryon}
\end{equation}
We thus find that when
\begin{equation}
  \frac{N_f}{N_c} > \frac{3}{11}(6-\sqrt{3}) \simeq 2.1,
\label{eq:bound-viol}
\end{equation}
all the bounds are violated---there is no light (protected, chiral) 
state that satisfies any of the bounds.  Note that the shift of $\Delta$ 
in Eq.~(\ref{eq:wgcads}) does not help because the dimensions of relevant 
operators are of $O(N_c) \gg 2$.  While it is logically possible that 
some unprotected operator satisfies a bound, we find it unlikely. 
Furthermore, even this loophole is closed when the theory is close 
to free, $N_f/N_c \simeq 3$.

A reasonable conclusion is that none of the inequalities in 
Eqs.~(\ref{eq:naive},~\ref{eq:wgcads},~\ref{eq:large-1},~\ref{eq:large-2}) 
applies universally, at least as are written.  This does not contradict 
the existence of the corresponding weak gravity bounds in the limit that 
theories admit weakly coupled gravity dual descriptions.  Even in the 
limit of large $N_c$, the supersymmetric QCDs considered here are expected 
not to have weakly coupled gravitational duals, as suggested e.g.\ by 
the presence of higher spin protected operators and a violation of the 
holographic central charge equality $c=a$.  The analysis here simply 
says that the bounds as are written cannot be true universally for 
all the CFTs.

The weak gravity bounds we present, therefore, must be corrected when 
we deviate from weakly coupled Einstein gravitational descriptions. 
In the example considered here, we are taking the large $N_c$ limit. 
Therefore, these corrections must be understood as higher derivative 
corrections (such as higher curvature terms).  It was claimed that 
higher derivative terms must contrive such that the original weak 
gravity conjecture holds without modification~\cite{Kats:2006xp}. 
(See also Ref.~\cite{Cremonini:2009ih} for related AdS discussions.) 
Our example suggests that this might not be the case in general.  Note, 
however, that the relation between the two analyses is not strict. 
For example, unlike in the Minkowski case, in AdS spacetime one cannot 
take a simple large black hole limit to make higher derivative terms 
be small perturbations to the system.

\subsection{CFT dual of extremal AdS/RN branes}
\label{subsec:branes}

Suppose the weak gravity bound from large black holes holds.  Then, 
any attempt to construct a dual field theory model for extremal 
AdS/RN branes must exhibit some instability, at least if we can take 
the weakly coupled limit in the gravity side.

There is, in fact, some attempt to construct field theories that 
model extremal AdS/RN branes in the large $N$ limit with long 
range interactions~\cite{Sachdev:2015efa}.  The claim is that it 
is possible to reproduce states with a large degeneracy matching 
with the Bekenstein-Hawking entropy of RN black branes.  An important 
thing for us is that these theories do not seem to show an instability 
suggested by the conjecture.

Similarly, in another recent paper~\cite{Hellerman:2015nra}, a universal 
behavior of scaling dimensions, $\Delta \sim Q^{(D-1)/(D-2)}$, in a large 
charge sector of certain (non-large $N$) $(D-1)$-dimensional CFTs was 
discussed.  The observation relevant to us is that the scaling behavior 
of $\Delta$ as a function of $Q$ is precisely that of large AdS-RN 
black holes in $D$ dimensions.  Again, as long as their effective 
field theory building on large charge expansion is valid, there does 
not seem any instability.

These analyses, however, do not immediately imply that the weak gravity 
bound from large black holes is invalid, since it is not clear if the 
theories analyzed have weakly coupled gravitational descriptions.  It 
would be interesting to study if these constructions can be applied in 
the regime in which the weakly coupled gravity limit can surely be taken. 
If such a limit can indeed be taken, the weak gravity conjecture for 
large black holes would imply that the effective field theory description 
discussed in Ref.~\cite{Hellerman:2015nra} must possess an additional 
instability mode.

\section{Discussion}
\label{sec:discuss}

In this paper, we have discussed possible generalizations of the weak 
gravity conjecture to AdS spacetime.  We have considered the conditions 
arising from both small and large AdS-RN black holes, and translated 
them into the language of dual CFTs.  While these conditions need 
to be satisfied a priori only in the regime in which weakly coupled 
gravitational descriptions are available, we have tested them against 
a wider range of CFTs.  We have found that the bounds as formulated in 
this paper are not universally satisfied by all the CFTs, and yet all 
the examples that we found do not satisfy them are theories that are 
expected not to have, or unclear to have, weakly coupled gravitational 
descriptions.

Although the bounds as written here do not apply universally to all 
the CFTs, it is possible that a similar, modified bound exists that 
is universally valid.  If such a bound exists, it must arise purely 
from consistency conditions applicable to all the CFTs with a $U(1)$ 
symmetry.  One candidate for such consistency conditions is the conformal 
bootstrap condition for correlation functions, and indeed there have 
been some studies on bounds of current central charges $C_V$ using 
this method (with or without fixing the dimensions of operators or 
the energy-momentum tensor central charge $C_T$)~\cite{Kos:2013tga}.

While these bounds given by the conformal bootstrap are rigorous within 
numerical precision, they mostly give lower bounds on $C_V$, yielding 
{\it lower} bounds on the strength of gravity.  Obtaining an upper 
bound is difficult because of the possibility that another (non-conserved) 
spin one operator mimics the current operator in question in a 
single bootstrap equation.  This makes it hard to isolate the relevant 
contribution.  In this respect, a promising case is a theory with 
$\mathcal{N}=2$ supersymmetry, in which a conserved current multiplet 
has an isolated contribution to the bootstrap equation so that the 
above problem can be avoided.  This allows us to obtain an upper bound 
on $C_V$~\cite{Beem:2014zpa}, although current theoretical technology 
still seems unable to extract a useful bound in this way for large 
values of $C_T$, which we are interested in.

Another direction would be to study the consistency of the CFT spectrum 
on nontrivial geometries such as $S_1 \times S_{d-1}$.  There is a 
constraint on the spectrum from the modular invariance in $d=2$ cases. 
There, the information of the central charges is also encoded in the 
torus partition function, and the promising results have been reported 
in Ref.~\cite{Hellerman:2009bu}.  In higher dimensions, however, due 
to the lack of manifest modular properties, it is an open question if 
we can derive an interesting bound.

\begin{acknowledgments}
We would like to thank Sean Hartnoll, Simeon Hellerman, Igor Klebanov, 
and Tomoki Ohtsuki for useful discussions.  Y.N.1 thanks Weizmann Institute 
of Science and Y.N.2 thanks Kavli Institute for the Physics and Mathematics 
of the Universe, University of Tokyo for hospitality during their visits 
in which a part of this work was carried out.  The work of Y.N.1 was 
supported in part by a Sherman Fairchild Senior Research Fellowship 
at California Institute of Technology and Department of Energy (DOE) 
grant number de-sc0011632 as well as the World Premier International 
Research Center Initiative (WPI Initiative), MEXT.  The work of Y.N.2 
was supported in part by the Director, Office of Science, Office 
of High Energy and Nuclear Physics, of the U.S.\ DOE under Contract 
DE-AC02-05CH11231, by the National Science Foundation under grants 
PHY-1214644 and PHY-1521446, and by MEXT KAKENHI Grant Number 15H05895.
\end{acknowledgments}

\appendix

\section{Normalization Convention}
\label{app:norm}

Here we present our normalization convention for $C_V$ and $C_T$ in 
Eqs.~(\ref{eq:def-CV},~\ref{eq:def-CT}).  In the dual gravitational 
description, our normalization for $C_V$ corresponds to taking that 
of $qe$ so that $C_D = (D-2)/(D-3)$ in Eq.~(\ref{eq:wgc-flat}).

We focus on $D=5$, i.e.\ 4-dimensional CFTs.  For a single complex 
scalar with a unit charge, we take its contribution to $C_V$ and $C_T$ 
as (see, e.g., Ref.~\cite{Barnes:2005bm}):
\begin{equation}
  C_V = \frac{1}{S_4^2},
\qquad
  C_T = \frac{8}{3} \frac{1}{S_4^2},
\end{equation}
where $S_4$ is the volume of the unit four-sphere, $S_4 = 2\pi^2$. 
For a single Weyl fermion with a unit (chiral) charge, we then have
\begin{equation}
  C_V = 2\frac{1}{S_4^2},
\qquad
  C_T  = 4 \frac{1}{S_4^2}.
\end{equation}
The contribution from a free massless vector field is $C_T = 16/S_4^2$.

The coefficient in the weak gravity bound, e.g.\ in the right-hand side 
of Eq.~(\ref{eq:naive}) can be worked out by noticing that the bound 
in the Minkowski limit (i.e.\ $1 \ll \Delta \ll C_T$) becomes identical 
to the BPS bound for the superconformal $R$-current in superconformal 
field theories, which is saturated by chiral primaries having $\Delta/q_R 
= 3/2$ and $C_T/C_R = 10$.

\section{Extremal AdS-RN Black Holes}
\label{app:AdS-RN}

In our normalization, the metric of the 5D AdS-RN black hole is given by
\begin{equation}
  ds^2 = -f(r) dt^2 + f^{-1}(r) dr^2 + r^2 d\Omega_3^2,
\label{eq:AdS-RN}
\end{equation}
where
\begin{equation}
  f(r) = 1 - \frac{2\kappa_5^2}{3r^2} M 
    + \frac{\kappa_5^2 e^2}{6r^4} Q^2 + \frac{r^2}{L^2},
\label{eq:f-r}
\end{equation}
and $M$ and $Q$ are the mass and charge of the black hole, respectively. 
The gauge potential is given by
\begin{equation}
  A_t = \mathrm{const.} - \frac{e^2 Q}{r^2}.
\label{eq:A_t}
\end{equation}
The outer horizon is located at $r= r_+$, where
\begin{equation}
  \frac{2 \kappa_5^2 M}{3} = r_+^2 
    + \frac{\kappa_5^2 e^2 Q^2}{6 r_+^2} + \frac{r_+^4}{L^2}.
\label{eq:outer-hor}
\end{equation}

The extremal limit is defined by
\begin{equation}
 Q^2 = \frac{6 r_+^4}{e^2 \kappa_5^2} \left( 1 + 2\frac{r_+^2}{L^2} \right),
\label{eq:extremal}
\end{equation}
so that $f(r)$ has a double zero with zero temperature.  In this limit, 
the mass of the black hole is given by
\begin{equation}
  M = \frac{3 r_+^2}{\kappa_5^2} 
    \left( 1+ \frac{3}{2}\frac{r^2_+}{L^2} \right).
\label{eq:M-extremal}
\end{equation}
Note that the BPS condition $M^2/Q^2 = C_D e^2 \kappa_D^{-2}|_{D=5} 
= (3/2) e^2 \kappa_5^2$ is not satisfied except for $r_+/L \to 0$, 
even though the black hole has zero temperature.

The transition between small and large black holes occurs at $r_+ \sim L$. 
At that point, $M \sim L^2 \kappa_5^{-2}$, and the corresponding conformal 
dimension is $\Delta \sim L^3 \kappa_5^{-2} \sim C_T$.

\end{document}